\def\grb{$\gamma$-ray burst \ }

\def\g{$\gamma$}
\def\G{\Gamma}
\def\etal{{\it etal. \ }}

\documentstyle[ichep,12pt]{article}
\title{\normalsize $\gamma$-RAY BURSTS AND NEUTRON STAR
MERGERS -
POSSIBLY THE STRONGEST EXPLOSIONS IN THE UNIVERSE\thanks{Supported by
NASA grant NAG5-1904}}
\author{Tsvi Piran\\
     Harvard-Smithsonian Center for Astrophysics,\\
     Cambridge, MA, 02138, USA\\
        and\\
     Racah Institute for Physics, The Hebrew University,\\
      Jerusalem, Israel, 91904}

\begin{document}
\finalcopy

\maketitle

\abstract{ $\gamma$-ray bursts have baffled theorists ever since
their accidental discovery at the sixties.  We suggest that these
bursts originate in merger of neutron star binaries, taking place at
cosmological distances.  These mergers release $\approx 10^{54}ergs$,
in what are possibly the strongest explosions in the Universe.  If
even a small fraction of this energy is channeled to an
electromagnetic signal it will be detected as a grbs.  We examine the
virtues and limitations of this model and compare it with the recent
Compton \g-ray observatory results.}

\vskip-1pc
\onehead{Prologue: $\gamma$-Ray Bursts circa 1973}

$\gamma$-ray bursts (grbs) were accidentally discovered ahead of their
time.  Had it not been for the need to verify the outer space treaty
of 1967 (which forbade nuclear experiments in space) we would not have
known about these bursts until well into the next century.  No one
would have proposed a satellite to look for such bursts, and had such
a proposal been made it would have surely turned down as too
speculative.  The VELA satellites with omni-directional detectors
sensitive to \g-ray pulses, which would have been emitted by a nuclear
explosion, were launched in the mid sixties to verify the outer space
treaty.  These satellites never detected any nuclear explosion.
However, as soon as the first satellite was launched it begun to
detect puzzling, perplexing and above all entirely
unexpected bursts.  The lag between the arrival time of the pulses to
different satellites gave a directional information
and  indicated  that
\break\newpage\vglue2pt\noindent
the  sources are outside the solar system.
Still, the bursts were kept secret for several years,
until Kelbsdal Strong and Olson described them in a seminal
paper\cite{Kle} in 1973.

The unexpected discovery sent theorist in a search for a source model.
Specific models ranged from comets to cosmic strings.

\onehead{False Clues?}

The rapid fluctuation in the signal (less than 10ms)
suggested a compact source, a neutron star or a black hole.
Several other clues focused the attention of theorists towards
neutron stars at the disk of the galaxy.

First, came an analytic estimate \cite{Sch} of the optical depth to
$\gamma \gamma \rightarrow e^+ e^-$.  For an impulsive source we have:
$\tau_{\gamma \gamma} \approx {\sigma_T F D^2 /R^2 m_e c^2}$
where F is the fluence ($\approx 10^{-5} ergs/cm^2$ in the
early detectors and $\approx 10^{-7}ergs/cm^2$ in
Compton-GRO), D is the distance to the source and $R$ its size
(expected to be less than $10^9$ from timing arguments).
Since $\tau_{\gamma \gamma} > 1$ for $D > 100 pc$, it
was argued that the sources must be at the disk of the galaxy.
Otherwise, it was argued an optical thick system will cool down and
radiate its energy in the x-ray uv or optical band and not as \g-rays.
The non-thermal spectrum also indicated that the sources are optically
thin. Incidentally,, it was the confrontation between this argument and
the indications from Compton-GRO that grbs are cosmological (which we
discuss later) that have lead to claims that grbs require ``new
physics". We will see that it ain't necessarily so.

A very strong and long (1000 sec) burst was observed on March 5th
1979.  The position of the burst coincided with a SNR remnant in the
LMC supporting  the idea that grbs originates on neutron stars.

Another clue came from the observation of absorption lines
\cite{Maz,Yosh}.  The lines were interpreted as cyclotron lines in a
$10^{12}$G magnetic field, a field strength that is found only on
neutron stars.

These clues and others have led to the consensus that grbs
arise  on neutron stars in the disk of the galaxy\cite{HL}, possibly
in their magnetosphere.

\onehead{Bursts distribution circa 1991}

There were, however some indications that the sources might not
be galactic.  In 1975 Usov and
Chibisov \cite{Uso} suggested to use a logN-LogS test to check if the
bursts have a cosmological origin.  Later, in 1983 van den Bergh
\cite{van} analyzed the distribution of the 46 bursts that were known
at that time and from the isotropy of this distribution he concluded
that the sources are either local at distances of less than half of
the galactic disk scale height or cosmological at redshift $z>0.1$
(See also \cite{Hb}).  The cosmological solution was accepted with
skepticism since with typical fluencies of $10^{-5} ergs/cm^2$ the
bursts require $10^{49} ergs$ if they originate at distances larger
than $100 Mpc$!  In 1986 Paczy\'nski \cite{Pac86} argued that the
bursts are cosmological and suggested that some of the burst are
lensed by intervening galaxies and that this will provide an
observational test to the cosmological hypothesis.  In 1989 Eichler,
Livio, Piran and Schramm\cite{Ei} (see also \cite{P90}) suggested that
the bursts originate in neutron star mergers at cosmological distances
(The possibility that grbs might be produced in neutron star mergers
was also mentioned without specifying a model by
\cite{Bel,Pac86,Goo,GDN}).  However, in 1991, just before the
Compton-GRO results were
announced, Atteia {\it etal.} \cite{Atteia} reported (at a $ 3\sigma$ level)
that the 244 bursts observed by the spacecrafts Venera 13, 14 and
Phebus  are concentrated towards the galactic plane, suggesting a disk
population after all.

\onehead{Neutron Star Binaries}

Another seemingly unrelated and unexpected discovery was make in 1975
by Hulse and Taylor \cite{Hul} who found a pulsar, PSR 1913+16, that
was orbiting around another neutron star.  No one has predicted that
such systems exist, but in retrospect it was not surprising.  More
than half of the stars are in binary systems. If some of these
binaries survive the two core collapses (and the supernovae
explosions) needed to produce the neutron stars they will end in a
binary pulsar.

The binary pulsar have proven to be an excellent laboratory for
testing General Relativity.  The binary system emits gravitational
radiation which is too
week to be detected directly, but its backreaction could be observed.
By carefully following the arrival time of the pulsar's signals Taylor
and his collaborator have measured the pulsar's orbit. They have
shown that  the binary spirals in
just in the right rate to compensate for the energy loss by
gravitational radiation emission (with two neutron stars, tidal
interactions and other energy losses  are
negligible).  For PSR 1913+16 the spiraling in takes place on a time
scale of $\tau_{GR} = 3 \times 10^8$years, in excellent agreement with
the general relativistic prediction \cite{Tay}.  These observations
not only confirm the general relativistic prediction, they  also
assure us that the orbit of the binary is indeed decreasing and that
inevitably in $3 \times 10^8$ years  the two neutron stars
will collide and merge!

\onehead{Source Count and Event Rate}

For many years only one binary pulsar was known. A simple estimate
based on the observation of one binary pulsar in several hundred
observed pulsars led Clark {\it etal.}  \cite{Clav} to conclude
that  about 1 in 300 pulsars is in a  binary. With a pulsars' birth
rate of one in fifty years this led to a binary birth rate of one in
$10^4$ years. Assuming a steady state, this is also the merger rate.
This estimate ignored, however, selection effects in the detection of
binary pulsars vs.  regular ones. Specifically PSR1913+16 is an
extremely bright pulsar which is detectable from much larger distance
than an average  pulsar. Currently there are four known binary pulsars
and an analysis based on their  luminosities  and life times
\cite{NPS,Phi} suggests that  there are $\sim 10^4 ~-~10^5$
neutron star binaries in the galaxy and that their merger rate is one
per $10^{6}$years per galaxy.  This corresponds to $\sim 100$ mergers
per year in galaxies out to a distance of 1 Gpc and about $10^3$ per
year to the horizon.  Narayan, Piran and Shemi \cite{NPS} also predict
that a similar or somewhat smaller population of neutron-star black
hole binaries will exist.

\onehead{Neutron Star Mergers}

It was immediately realized, after the discovery of PSR1913+16 that
the binary produces a  unique chirping gravitational radiation signal
during the last seconds  before the neutron stars merge.  These
signals are probably the best candidates for detection of
gravitational radiation.  However, these events are  rare and to
observe them (gravitationally) in our life time we must turn to
extragalactic events.  This is the aim of the advanced gravitational
radiation detectors like LIGO \cite{LIGO}.

As the strongest sources of gravitational radiation neutron star
mergers attracted the attention of relativists, but most astronomers
ignored them as being too rare to be of interest. Clark and Eardley
\cite{Clae} have shown that the binding energy released in a neutron
star binary merger is $\sim5 \times 10^{53} ~-~10^{54}ergs$, making
these events possibly the most powerful explosions in the Universe. A
significant fraction of this energy is emitted as gravitational
radiation, both prior and during the collision.  A very sophisticated
gravitational radiation detector, LIGO, is built to detect these
gravitational radiation signals. But it will be around the turn of the
century when it is operational.

As the neutron stars collide a shock forms and the stars heat up.
Most of the binding energy is emitted as neutrinos \cite{Clae}. The
neutrino burst is comparable or slightly stronger than a supernova
neutrino burst (such as the one detected by Kamiokande and IMB from
1987A). To detect extragalactic events at cosmological distances we
need a detector which is $\approx 10^8$ times larger than those
detectors. With regular supernova neutrino bursts being a hundred
times more frequent it is clear that these neutrino signals are not
the prime candidates for detection.

Neutron star
mergers are hiding from us by emitting their energy in two channels
with extremely small cross sections.  If even a small fraction of the
energy is channeled to an electromagnetic signal, its much large cross
section will make it much easier to observe. For many years, I kept
wondering what are the possible observational consequences of such
events \cite{P90}.

\onehead{Energy Conversion}

Goodman, Dar and Nussinov \cite{GDN} suggested that the neutrino-anti
neutrino annihilation $\nu + \bar \nu \rightarrow e^+ + e^-$ converts a
small fraction of the neutrino supernova burst to electron-positron
pairs which in turn annihilate to \g-rays, heat the surrounding
envelop and provide the energy required to power the supernova shock
wave.  In 1989, Eichler, Livio, Piran and Schramm \cite{Ei} (see also
\cite{P90}) suggested that the same mechanism operates in neutron star
mergers and converts $\sim10^{-3}$ of the emitted energy to pairs and
\g-rays.  This corresponds to $10^{51} ergs$, roughly sufficient for
detection of the bursts from cosmological distances. Eichler {\it
etal.} \cite{Ei} used the old estimate of Clark {\it etal.}
\cite{Clav} for the merger rate and suggested that these events would
be detected by Compton observatory  as grbs.

More recently, alternative energy generation mechanism such as
magnetic field recombination \cite{NPP} or accretion onto the neutron
star \cite{Pac92a} have been proposed and it was argued that they
provide comparable amounts of energy.

\onehead{Fireballs and Relativistic Effects}

If grbs are indeed cosmological they are initially optically thick,
as Schmidt \cite{Sch} have argued. How can there be a \grb from such a
source?  Goodman \cite{Goo} considered a dense sphere of \g-ray photons
and pairs, which he called a fireball.  He has shown that as long as
the fireball is optically thick the radiation-pair plasma will behave
like a fluid with $p = \rho/3$. The fireball will expand and cool,
just like the early Universe (unlike our Universe the gravitational
force is unimportant). As the fireball cools  its temperature drops
with $T \propto 1/R$ until the electron positions annihilate (the
annihilation is complete at $T \approx 20$  keV) and the radiation
escapes. The radiation fluid has reached in the meantime a
relativistic velocity relative to an observer at infinity and its
Lorentz factor $\Gamma \approx R_{esc}/R_0 \approx T_0/T_{esc} \approx
10^3 - 10^4$. The escaping photons, which have a typical energy of 20
keV in the local frame are blue shifted relative to an observer at
infinity and their observed energy is $\epsilon_{obs} \approx \G
T_{esc} \approx T_0$, of the same order as the initial energy. In this
way the optical depth argument which limited the distances to the
sources is bypassed and  there is no need to introduce ``new
physics" to explain grbs from cosmological distances.

Paczy\'nski \cite{Pac86} have shown that similar effects take place if
the radiation is released in a quasi-stationary manner. In this case
the radiation flows out as a relativistic wind, with $T\propto 1/R$
and $\Gamma \propto R$. The radiation ceases to behave like a fluid
and escapes when $T \approx 20$ keV in the local frame. The escaping
x-ray photons are blueshifted to much higher energies in the observer
frame.

\onehead{Do Fireballs Work?}

The fireball model faces two serious objections: the origin of the
observed nonthermal spectrum and the effects of baryons.

There is no clear way to explain the non-thermal spectrum from a
fireball that passes an optically thick phase and termalizes.  It
is possible that different regions in a realistic, inhomogeneous
fireball move with significantly different Lorentz $\Gamma $ factors
and that the observed spectrum is a blending of thermal spectra to a
non thermal one. Simple calculations of the spectrum of a spherical
fireball \cite{SPN} show some deviation from a thermal spectrum, but
it is not large enough.  Alternatively one could hope that the
spectrum would become nonthermal in the transition from optically
thick to optically thin regimes.  However, this transition takes place
at $\approx 20$keV in the local frame. The energy injected from
annihilation at this stage is insignificant and the temperature is too
low for inverse Compton scattering to be effective \cite{NPY}.  It
seems that there is no clear mechanism that will modify the photons'
black body spectrum in this stage.

One expects that some baryons will be ejected into the fireball.
Shemi and Piran \cite{SP} have shown that the baryons have two
effects.  For $10^{-11}M_\odot<M<10^{-8}M_\odot (E_0/10^{51}ergs)$ the
baryons dominate the opacity (long after all the pairs have
annihilated) without influencing the fireball's inertia. The fireball
continues to be optically thick until $\tau_g = \sigma_T M / R^2 = 1$.
This leads to a longer acceleration phase and to a larger final
Lorentz factor $\Gamma_f \approx R/R_0\approx T_0/T$.  However, the
final energy of the escaping radiation remains unchanged with
$\epsilon \approx \Gamma T \approx T_0$.

Larger baryonic load changes the dynamics of the fireball.  As the
fireball expands $\rho \propto R^{-3}$ while $e \propto r^{-4}$. If $M
> 10^{-8}M_\odot (E_0/10^{51}ergs)$ the baryonic rest mass will
dominate the energy density and the fireball's inertia before the
fireball becomes optically thin.  In these cases all the energy will
be used to accelerate the baryons with $E_K = Mc^2 \Gamma \approx (E_0
+ Mc^2) / (E_0 T / Mc^2 T_0 + 1)$.  The final outcome of a loaded
fireball will be relativistic expanding baryons with $\Gamma \approx
E_0 / M c^2$ and no radiation at all.

Several ideas have been proposed to avoid the baryonic load problem.
These include: (i) Separation of the radiation and the baryons due to
deviations from spherical symmetry - the radiation escaping along the
axis and the baryons being ejected preferably in the equatorial plane
\cite{NPS} and (ii) generation of a radiation fireball with very
small amounts of matter  via magnetic processes \cite{NPP}.

\onehead{Energy Conversion, Once More}

If the baryonic contamination is in the range $ 10^{-5} E_0<M c^2 <
0.1 E_0$ all the initial fireball energy will be converted to
extremely relativistic protons moving at a  Lorentz
factors $10< \Gamma \approx E_0 / M c^2< 10^5$.  Recently
M\'es\'zaros, and  Rees \cite{MR} (see also \cite{NPP}) suggested that
this energy could be converted back to \g-rays when this baryons
interact with the surrounding interstellar matter.  A shock, quite
similar to a SNR shock,  forms and it cools predominantly via
synchrotron emission in the x-rays.  The  x-ray photons will be
blueshifted to \g-rays in the observer frame due to the  relativistic
velocity of the fireball.  The relativistic motion will also lead to a
short time scale for the burst.  Alternatively, the accelerated
baryons could interact with a pre-merger wind that surrounds the
fireball \cite{NPP,NPY}.  In both cases the interaction with the
surrounding material will lead once more to the conversion of the
energy: from kinetic energy back to radiation.  Since this phase is
taking place in an optically thin region the  photons will  not
thermalize and the emerging spectra will be non thermal, as observed.
Thus, this process seems to resolve at one stroke
both major objections to the fireball scenario at one stroke.

\onehead{\g-ray bursts distribution circa 1992}

The Compton \g-ray observatory was launched in the spring of 1991
(see \cite{Paci} for a review).  It
includes an omnidirectional \grb detector (BATSE) which, with a
limiting sensitivity of $\approx 10^{-7} ergs/cm^2$, is the most
sensitive detector of this kind flown.  By the summer of 1992  BATSE has
detected more than 400 bursts, more than all previous detectors
combined.  BATSE is also capable of obtaining a directional information
on the bursts on its own.
Within four month from its launch BATSE has collected
enough data to conclude that the distribution of grbs sources is
isotropic \cite{Mee}.
When the $V/V_{max}$ test was applied to the burst intensity distribution
is was shown that the sources are not distributed homogeneously in
space and that there is a concentration of sources towards us \cite{Mee}.

These two observations rule out all local galactic disk models.
The observations are consistent with three possible populations:
(i) Cosmological population (ii) Galactic halo population with a large
core radius ($>50kpc$) and (iii) A population, such as comets at the
Oort cloud, centered around the solar system.

We will turn to the second and third possibilities, before summing up
the status of the cosmological population models.

\onehead{Galactic Halo models}

Galactic Halo models require a halo population with a large core
radius (to avoid an anisotropic enhancement towards the galactic
center). This is a new population of astronomical objects, which was
not seen elsewhere \cite{Pac92a}.  It probably require a different
distribution (in space) than the dark halo material (the latter being
too concentrated towards the galactic center).  By now there have been
several suggestions how to form a neutron population of this kind.
These include either ejection from the galactic disk  or formation
in site. However, these models face additional difficulties.

Approximately $10^{41} ergs$ are needed for bursts at the halo, quite
a large amount for a neutron star.  With a typical size of $10^6cm$
this leads to an optical thickness of $\approx 10^8$ for $\gamma
\gamma \rightarrow e^+ e^-$.  These constraints have two far reaching
implications: First, the optically thin neutron star models suggested
for galactic disk sources are inapplicable to grbs at the halo.
Second, galactic halo sources inevitably involve an opaque pair plasma
fireball, just like cosmological sources \cite{PS2}.  The physical
conditions in these fireballs are, however,
less favorable than the conditions
at cosmological fireballs for production of grbs.

\onehead{Local Population}

Typical objects in the solar system have a very small binding energy per
baryon and it is difficult to imagine a mechanism in which such
objects generate energies in the \g-ray range (see however
\cite{Katz}).  The only hope is probably via a magnetic phenomenon.
Solar flares do generate grbs which are detected by Compton-GRO (these
are identified by their location and spectrum \cite{Fish}). However,
comparison of the size and masses involved in these events make it
inconceivable that similar conditions can be achieved elsewhere in the
vicinity of the solar system, without leaving any other trace.

\onehead{Cosmological Population}

Several groups \cite{MP,P92,Der,Schm} have shown that a cosmological
population is compatible with the observed $V/V_{max}$ distribution.
The apparent concentration towards us is an artifact of a combination
of redshift effects and a possible cosmological evolution.
Unfortunately it is impossible to separate the two effects and to
determine the typical red shift, $\langle Z \rangle $, to the sources
from the $V/V_{max}$ distribution.  Depending of the cosmological
model and the source evolution we have $0.3<  Z_{av}  < 3 $.
For $\Omega=1$ and no evolution  $  Z_{av}   \approx 1$
\cite{P92}.

The event rate needed to explain the
observation is in an amazing agreement with the rates estimated for
neutron star mergers \cite{NPS,Phi}.  Because of a historical  coincidence
the forth binary pulsar, PSR1534+12, which played a decisive role in
the determination of the merger rate \cite{NPS,Phi}, was discovered
\cite{Wol} a few month before Compton-GRO was launched and the
prediction of the neutron star merger rate were not influenced by the
rates required to explain the Compton-GRO results.

Several other
cosmological grbs models were suggested after Compton-GRO
\cite{Car,McB,Hoy,Uso92}. Within the cosmological framework,
the neutron star merger scenario is the most conservative one
possible.  It is the only one based on a source population that
definitely exists. We know its members will merge, we can be certain
that huge quantities of energy will be released in such mergers, and
we find the merger rate to be comparable to the observed burst rate.

\onehead{Clues Revisited}

Before concluding we turn once more to the  clues
discussed earlier. The optical depth problem disappeared in some sense
and remained in another.  Relativistic effects, due to the expansion of
the fireball \cite{Goo,Pac86}, were not taken into account in the
original argument \cite{Sch} which is flawed. The resulting spectrum
from the expanding fireball has the right energy range but to a first
approximation it is thermal.  It is a non-trivial (but not impossible)
task to obtain a nonthermal spectrum.  This problem is shared by all
cosmological and galactic halo models.

The March 5th event was one of three soft \g-ray repeaters, which have
a softer spectrum and produce  repeated bursts from the same source,
unlike all other sources \cite{HL}.  It is by now generally accepted that these
are most likely a different phenomenon.

The nature of the cyclotron lines has been fairly controversial since
they were first reported\cite{Lar,Hard}.  Mazets {\it etal.}
\cite{Maz} claimed that single ``cyclotron absorption lines" were
present in 20 bursts, with a broad distribution of line energies
(27--70 keV), but with only five lines having energies under 50 keV.
This is in conflict with the GINGA experiment which discovered three
systems of lines, all with nearly identical energies, all under 50 keV
\cite{Yosh}.  So far, no lines have been detected with any experiment
on the Compton Gamma Ray Observatory.

\onehead{Epilogue: \g-ray bursts circa 2000}

A general test for all cosmological models is the expected positive
correlation between the faintness of a burst (correlated with
distance) and redshift signatures through the burst duration and
spectrum \cite{Pac92b,P92}.  This correlation could be masked by large
intrinsic variations among bursts, but should eventually be observed
when enough data accumulate.

At present there are no known optical counterparts to grbs.  Since
neutron star binaries might be ejected from dwarf galaxies,  we predict
\cite{NPP}, that
grbs occur within a few tens of arcsecond
from dwarf galaxies and within but not necessarily at the center of
ellipticals. Optical identification of some parent galaxies, could
support this model and the location of the burst relative to the
galaxy could distinguish this model from other cosmological scenarios
that involve supermassive black holes or other objects located in the
centers of galaxies \cite{Car,McB,Hoy}.

The scenario makes one unique prediction: strong $\gamma$-ray bursts
should be accompanied by a gravitational wave signal \cite{PNS,P92,NPP}
(though the reverse need not necessarily be true if the $\gamma$-rays
are beamed).  These signals should be detected by LIGO \cite{LIGO}
when it becomes operational (hopefully by the year 2000).  LIGO should
provide good distance estimates to individual bursts \cite{Schu} and
should also pinpoint the exact time of the merger, in addition to
furnishing an ultimate proof of this model

\def\ApJ{{\it Ap. J.}}
\def\ApJL{{\it Ap. J. L.}}
\def\Nature{{\it Nature}}

\end{document}